\documentstyle[preprint, eqsecnum, aps, epsfig]{revtex}
\tightenlines
\newcommand{\omits}[1]{}
\begin{document}

\title{Difference Discrete Variational Principle\\
 in Discrete Mechanics and Symplectic Algorithm}

\author{Xu-Dong Luo
\footnote{Email: luoxd@sjtu.edu.cn }}
\address
{Department of Physics, Shanghai Jiao Tong University, \\[0pt]
Shanghai 200030, China}
\author{Han-Ying Guo, $\quad$ Yu-Qi Li
\footnote{Email: hyguo@itp.ac.cn ,$\;$ qylee@itp.ac.cn}}
\address
{Institute of Theoretical Physics, Chinese Academia of Sciences, \\[0pt]
P.O. Box 2735, Beijing 100080, China}
\author{Ke Wu
\footnote{Email: wuke@itp.ac.cn } }
\address
{Department of Mathematics, Capital Normal University,\\[0pt]
 Beijing 100037, China}

\maketitle
\begin{abstract}
We propose the difference discrete variational principle in
discrete mechanics and symplectic algorithm with variable
step-length of time in finite duration based upon a noncommutative
differential calculus established in this paper. This approach
keeps both symplicticity and energy conservation discretely.
\omits{ on high order jet bundle, then deal with variation
problems in discrete mechanics from the viewpoint of NCDC
completely.} We show that there exists the discrete version of the
Euler-Lagrange cohomology in these discrete systems. We also
discuss the solution existence in finite time-length and its site
density in continuous limit, and apply our approach \omits{the
obtained algorithms }to the pendulum with periodic perturbation.
The numerical results are satisfactory.

\end{abstract}

\newpage
\tableofcontents
\newpage


\section{Introduction}
In the course of numerical calculations\omits{, motion equations }
in continuous mechanics,  the equations of motion\omits{ may be
difficult to solve in general, so that we have to} should be
discretized in time variable to get certain \omits{ the motion
equations and get its approximations or }schemes. \omits{Since
there have many kinds of discretization methods, it is natural to
ask which one has priority. For example,}Among  various discrete
time schemes in Hamiltonian mechanics, the symplectic algorithm is
quite successful \cite{ruth83,fk84,sc94}, since it preserves the
symplecticity.\omits{ shows its power in long time calculation, so
we believe the symplectic-preserving discretization on motion
equations is better in Hamiltonian mechanics.}
In Lagrange mechanics,  the  discretizations may be established
via discrete variation method to get the symplectic-preserving
integrators. But this variation is usually taken with respect to
the discrete dependent variables only \cite{td1},\cite{av88}.

Thus, for the long time, there had been no discrete Legendre
transformation. Therefore, the discrete Lagrange and Hamilton
mechanics could not be transformed to each other. In addition, the
equal-time-step symplectic integrators \cite{av88,MV91} cannot
preserve the energy, while the variable time-step \omits{. There
have some discrete versions of variational principle} variational
integrator proposed in \cite{td1,td2} preserves discretely energy
but without proving the symplicticity. \omits{While in
\cite{av88,MV91}, the schemes with fixed time-step are symplectic
so that the energy conservation is lost.} \omits{in discrete
Lagrangian mechanics, emphasizing particularly on the conservation
of first integral and differential structure, respectively.}Until
recently, it is proposed \cite{kmo99,cgw,cgw1} that the energy
conservation and symplicticity for the variational integrators
with variable time-steps may be compatible. However, it is still
open how to deal with discrete Hamilton systems as well as
symplectic algorithm by variational principle. In the meantime,
the difference discrete variational principle (DDVP) has been
introduced in \cite{glw,glww}. And in this approach the discrete
Legendre transformation can be introduced and both discrete
Lagrange systems including variational integrators and Hamilton
systems including the schemes in symplectic algorithm can be
transferred to each other. One of the key points in DDVP is that
the difference is treated as an independent variational variable.
Very recently, DDVP with variable time step (VDDVP) has been
proposed in \cite{gw}. Thus the symplicticity and energy could be
discretely preserved at equal footing.

In this paper, we propose the non-commutative differential
calculus (NCDC) on fibre  bundle over discrete base space in order
to deal with the variation of some Lagrange system with high order
derivatives. Based upon this NCDC we present  the VDDVP
with  variable time steps as well as the discrete version of the
Euler-Lagrange cohomology and apply them to discrete mechanics
including the generalized symplectic algorithm. \omits{ Let $d_t$
be the total exterior differential operator,The main content may
be enumerated as follows: First,}For a given discrete action $S$
within finite time-duration, we obtain both discrete equations of
motion and energy conservation equation from the total variation
$\delta S:=i_{\xi_{total}} d_t S =0$, where $d_t$ is the total
exterior differential operator, by means of VDDVP with fixed end
points of the duration. \omits{This leads to both the discrete
equations of motion and energy.}Then we show that the discrete
version of the first Euler-Lagrange cohomology does exist in the
case of variable time-step. We also consider some simple example
to show the advantages of our approach.

\omits{In  discrete mechanics of variable
step-lengths of discrete time we propose the
discrete variational principle provided the
Lagrangian is dependent on time explicitly and
high order time derivative. In order to do that ,
we establish a class of noncommutative
differential calculus (NCDC) on high order jet
bundle, then deal with variation problems in
discrete mechanics from the viewpoint of NCDC
completely. It shows the nontrivial
Euler-Lagrange cohomology in discrete mechanics
exist.}



The paper is organized as follows. In section 2, we propose an
NCDC on the fibre  bundle whose base space is 1-dimensional
lattice with variable step-lengths. In section 3, we present the
VDDVP and deal with both discrete Lagrangian and Hamiltonian
mechanics based upon such an NCDC. In section 4, we discuss the
solution existence in finite time step-length and site density in
continuous limit. In section 5, by means of the VDDVP, we obtain
energy-conserving midpoint scheme for Hamiltonian system that may
depend on time manifestly and show its numerical properties.
Finally, we end with some remarks and discussions.

\section{The NCDC on Lattice with Variable Spacing}

\omits{Instead of ordinary differential calculus on smooth
manifold,}For the lattice with equal spacing, an NCDC on its
function space 
has been proposed in \cite{gwz}. In this section, we generalize it
to the one on the lattice with variable spacing\omits{high order
jet bundle} in order to describe the VDDVP on discrete mechanics
including energy-preserving and symplectic algorithm.

\subsection{The NCDC on Lattice with
Variable Spacing}

For the simplicity, we consider one dimensional case. It
corresponds to the time variable $t\in R$  and is dicretized as
$t_k\in T_D$,
\begin{equation}\label{td0} t\in R  \rightarrow
t\in  {T_D}=\{ (t_k , t_{k+1}=t_k+\Delta t_k,
\quad k \in I)\}. \end{equation}
 Note that $\Delta t_k$ are not fixed here. The total number
of nodes in set $I$ is finite for the case of finite
interval
, while it is is infinite if the interval
 is infinite. We
will discuss  both of them. The algebra of all the functions
defined on $T_D$ is denoted as $\cal A$. There is an algebra
homomorphism defined by the shift operator as follows
 \begin{equation}
       R_E:  {\cal A} \rightarrow {\cal A},\quad\quad
         R_E f(t_k)=f(t_{k+1}).
\end{equation} Then one introduces the vector
fields $V=span\{\partial_E\}$ on
 $\cal A$, where
 \begin{equation}\label{drv}
(\partial_E f)(t_k):= (R_E-id) f(t_k) = f(t_{k+1})-f(t_k).
\end{equation}
 The space of
1-form $\Omega ^1(T_D)=span \{ \chi^E\}$ is dual to $V$:
$\chi^E(\partial_E)=1 $. The whole differential algebra
$\Omega^*(T_D)$ can  be defined as
$\Omega^*(T_D)\bigoplus\limits_{n=0,1} \Omega^n(T_D) $ with ${\cal
A}=\Omega^0(T_D)$. Introduce the exterior differentiation in
$\Omega^*$ $ d: \Omega^0(T_D) \rightarrow \Omega^{1}(T_D)$. It
acts on a $0$-form $f \in \Omega^0 (T_D)= \cal A $ gives
\begin{equation}\label{df}
 df:=(\partial_E f) \chi^E \quad
\in \Omega^{1}. \end{equation} An NCDC with respect to $d$ on
$\cal A$ is constructed \cite{gwz}. As a simplified version of
\cite{AC94}, its nilpotency and Leibniz law of its acting on forms
lead to the noncommutative property  in $\chi^E
 f(a)=(R_E f(a)) \chi^E $.

It is convenient to consider a set of auxiliary
bases of $\Omega^1$: \begin{equation}\label{base}
 dt_k := (t_{k+1}-t_k) \; \chi^E .
\end{equation} Thus the exterior differentiation
(\ref{df}) becomes $df(\cdot)= [\Delta_k
f(\cdot)] \;dt_k$, where $f\in \cal
A\omits{^{\prime}}$\omits{, the function space on
lattice $L^1$,} and $\Delta_k
f(\cdot):=(t_{k+1}-t_k)^{-1} (R_E-id)\;
f(\cdot)$.

For each interval $[t_k, t_{k+1}]$ on lattice
$T_D$, the discrete integral over it can be
defined as follows: At moment $t_k$, if there has
a $1$-form $\Delta_k f(t_k)\;dt_k$, whose
integral over the path $t_k\rightarrow t_{k+1}$
is \begin{equation}\label{itg}
\displaystyle\int_{D, t_k}^{t_{k+1}} \Delta_k
f(t_k)\;dt_k :=\displaystyle\int_{D,
t_k}^{t_{k+1}}{\cal D}t_k\; \Delta_k f(t_k) f(t_{k+1})-f(t_k), \end{equation} where
$\displaystyle\int_{D}$ denotes discrete
integral, and ${\cal D}t_k$ is the measure at
moment $t_k$. Under this definition\omits{In this
way}, any integral over the path $t_i\rightarrow
t_{j}$, $i\leq j$, can be divided into $(j-i)$
parts:
\begin{equation}\begin{array}{rcl}\label{stoke}
\displaystyle\int_{D,
t_i}^{t_j}df(t)&=&\displaystyle\sum_{i\leq k
\leq j-1}\int_{D, t_k}^{t_{k+1}} [\Delta_k f(t_k)]\;dt_k \\
  &=&\displaystyle\sum_{i \leq k
\leq j-1} (f(t_{k+1})-f(t_k))
  =f(t_{j})-f(t_{i}).
\end{array} \end{equation} Here discrete integral
(\ref{itg}) can be regarded as the analogues of the ordinary
integral in continuous case, and the equation (\ref{stoke}) is the
discrete version of Stokes' formula.

\subsection{The NCDC on Fibre Bundle over Discrete Space}

In order to establish the differential calculus on fibre bundle
over discrete space, we \omits{first introduce the concept of
fibre bundle with discrete base space. C}consider 
the base space, time $t$, is
 discretized as
\begin{equation}\label{td} t\in R  \rightarrow
t\in  {T_D}=\{ (t_k , t_{k+1}=t_k+\Delta t_k,
\quad k \in I)\}, \end{equation} but its
$n$-dimensional fibre space $M_k$, at moment $t_k
(k \in I)$, is still continuous and smooth
enough. Let $N$ be the set of all nodes on
${T_D}$ with index set $Ind({N})=I$, and
${M}=\bigcup_{k \in I} M_k$ be the fibre space on
$T_D$.
 At the moment $t_k$, ${\cal
N}_k$  denotes the set of nodes neighboring to $t_k$, and ${I}_k $
is the index set of nodes of ${\cal N}_k$ including $t_k$. The
coordinates of $M_k$ are denoted by $q^i({t_k})=q^{i (k)}, i=1,
\cdots, n$.  Let ${\cal M}_k=\bigcup_{{l} \in {I}_k} M_l $ be the
union of fibre spaces $M_l$, ${l}\in {I}_k$ on ${\cal N}_k$, and
$F({\cal M}_k)$ be the function space.

The  exterior differential calculus on $F({\cal M}_k)$ consists of
two parts: the vertical differentiation $d_v$ along the fibre
spaces and the horizontal differentiation $d_h$ along the base
space \cite{gwz}. Note that the operator $d_v$ acts on fibre
variables only, while the operator $d_h$ also acts on function
space $F(T_D)$. Obviously, on the function space $F(T_D)$, the
operator $d_h$ should just play the role of operator $d$, which is
already established in \cite{gwz}. Thus, at moment $t_k$, on total
space ${\cal M}_k \times T_D$, the total differential operator on
function space $F({\cal M}_k\times T_D)$  locally is $d_t
:=d_v+d_h$. When the functional Lagrangian under consideration
depends on time $t$ manifestly, we should consider that $d_h$ acts
not only on $F({\cal M}_k)$, as the discrete differentiation with
respect to $t_k$ via the variables on fibre spaces ${\cal M}_k$,
but also on $F(T_D)$ as the differentiation directly for variables
$t_k$.

 For the local coordinates
$q^i$ on the fibre, there are
\begin{equation}\label{dtq} \begin{array}{rcl}
 \quad d_t q^{i(k)} &:=& d_v q^{i(k)} + d_h q^{i(k)},\\
       d_h q^{i(k)} &:=& (R_E\;q^{i(k)}-q^{i(k)})\;\chi^E
              = (q^{i(k+1)}-q^{i(k)})\;\chi^E,
\end{array}
\end{equation}
where $R_E$ is translation operator on the fibre bundle space. For
the  $f\in F({\cal M}_k\times T_D)$, the operator
  $R_f$ acting only on fibre
spaces is also needed.   They are defined as follows,
\begin{equation}\begin{array}{rcl}\label{rerf}
R_E \; f(\{q\};\{t_k\})&:=& f(\{R_E\;q\};\{R_E \;t_k\}), \\
R_f \; f(\{q\};\{t_k\})&:=&
f(\{R_E\;q\};\{t_k\}). \end{array}\end{equation}
The symbol $\{f\}$ denotes  the set of variables
of same kind with the function $f$. For example,
$\{q\}$ is the set of variables on fibre space,
and $\{t_k\}$ is the set of variables on base
space, etc.

Then we can define the differential calculus on
fibre bundle over discrete base space,

{\it {\bf  Definition 2.1:} On the total space ${\cal M}_k \times
T_D$, the NCDC on function $f\in F({\cal M}_k \times T_D)$ is
given by : \begin{equation}\label{dtf} d_t f(\{q\};\{t_k\})
:=\displaystyle \sum_{i,k\in\{q\}} (\partial_{q^{i(k)}} f)\; d_t
q^{i(k)}
     + ( R_E \;f - R_{f}\; f)\; \chi^E .
\end{equation}}

Now, let us construct the higher order differential algebra. As is
well known,  $d_v$ is an ordinary differential operator satisfying
the ordinary Leibniz rule and acting on $F(M)$ only. Its
differential complex is also as usual and denoted as $\Omega^*(M)\bigoplus\limits_{n} \Omega^n(M) $. The differential complex for
$d_t$ is the topological  product of $\Omega^*(M)$ and
$\Omega(T_D)$. The differential complex of $\Omega(T_D)$ is
noncommutative and with only one generator
 $\chi^E$. Any functions and $\Omega(T_D)$ is not commutative.
Acting $d_v$ on the
noncommutative relation
$\chi^E\;q^{i(k)}=q^{i(k+1)}\;\chi^E$, there is $
-\chi^E\wedge d_v q^{i(k)} =  d_v
q^{i(k+1)}\wedge \chi^E$. It can be generalized
to any differential algebra $\omega$ on the fibre
bundle: \begin{equation}\label{wedge}
  \chi^E\wedge\omega = (-1)^{deg(\omega)}\;(R_E\;\omega)\wedge \chi^E .
\end{equation}
   The operation of total operator $d_t$ on $d_v q$ is defined
as follows
\begin{equation}\begin{array}{rcl}\label{rule1}
d_v \: d_vq^{i(k)}&:=& 0 ,\\
 d_h \;d_vq^{i(k)}&:=&-(R_E\; d_vq^{i(k)}-d_vq^{i(k)})\wedge
\chi^E.\\
\end{array}\end{equation} It is obvious to get
that $d_v \chi^E=0$ and  ${ d_h}^2=0$ since there
is only one 1-form on base space and it is
antisymmetric. We can get,
\begin{equation}\begin{array}{rcl}
d_v \; d_h q^{i(k)}&=&(d_vq^{i(k+1)}-d_vq^{i(k)})\wedge \chi^E \\
     &=& - d_h \;d_vq^{i(k)}.
\end{array}\end{equation}

Thus, for the local coordinates $q^i$ of fibre,
 a well-defined differential
calculus  has been established and the total differential operator
$d_t=d_v+ d_h$ is indeed nilpotent:
\begin{equation}\begin{array}{rcl}
  d_v^2&=&  d_h^2 =0,\\
  d_t^2&=& (d_v+ d_h)^2 =0.
\end{array}\end{equation}

\subsection{The NCDC on Extended  Bundle over Discrete Space}

In \cite{av88}, \cite{MV91}, Veselov uses
$Q\times Q$ for the discrete version of the
tangent bundle of a configuration space $Q$. The
Lagrangian $L(q,\dot{q})$ is discretized to
$L(q^{(k)},q^{(k+1)})$. In his approach, how to
introduce the discrete Legendre transformation
and discrete canonical momentum had not been
considered. In addition, the time step-lengths
are fixed with equal step-length. Very recently
two of us with their collaborators in \cite{glw},
\cite{glww},\cite{gw} proposed another discrete
variational principle and introduced discrete
Legendre transformation and discrete canonical
momentum. In terms of  NCDC on the time lattice
(the base space) with variable step-lengths,
exterior differentiation rules of variables $t_k$
's and $q^{i(k)}$ 's are taken on different ways.
We will follow this line to discuss the discrete
variational principle with high order derivatives
or differences.

 The
continuous Lagrangian
$L(q^{i},\dot{q}^{i},\cdots,(\frac{d}{dt})^{l}q^{i};t)$ can be
discretized   in such a form
$L^{(k)}:=L(v_0^{i(k)},v_1^{i(k)},\cdots,v_l^{i(k)};t_k)$ where
 $v_0^{i(k)}:=q^{i(k)}, v_1^{i(k)}:=\Delta_k q^{i(k)}$, $ v_m^{i(k)}:=\Delta_k
v_{m-1}^{i(k)}, 2\leq m\leq l$. \omits{In  continuous variational
principle with high order derivatives the terminology of jet
bundle are widely used. We shill establish the concept of jet
bundle over discrete one dimensional base space. For fibre bundle
over discrete space $E(M, T_D)$ with coordinate $(q^{i(k)},t_k)$,
its first jet extension is denoted as $J^{(1)}(E(M, T_D))$ with
local coordinates $J^{(1)}_k=(q^{i(k)},v_1^{i(k)};t_k)$. At the
section of  $q^{i(k)}(t_k)$, the extended coordinates $v_1^{i(k)}$
is $v_1^{i(k)}:=\Delta_k q^{i(k)}$.} In order to deal with the
variation problem of the Lagrangian with $l-th$  order
derivatives, we first consider that the all variables
$(v_0^{i(k)}, v_1^{i(k)}, \cdots,v_l^{i(k)})$ are independent from
each other and denote the fibre bundle over one dimensional
discrete space $T_D$ as $\overline{E}(\overline{M}, T_D)$ , then
use the constrains to fix the relation between them
$v_0^{i(k)}:=q^{i(k)}$, and  $ v_m^{i(k)}:=\Delta_k
v_{m-1}^{i(k)}, 1\leq m\leq l$.

At each point of $t_k\in T_D$ the fibre space $\overline{M_k}$ is
of dimension $n \times (l+1)$ and with local coordinates
$(v_0^{i(k)}, v_1^{i(k)}, \cdots,v_l^{i(k)})$. In order to define
 the exterior derivatives  we should discuss the functions defined on
 the nodes neighboring to $t_k$. Similar to the discussion in the
 last subsection we
 can define ${\overline{\cal
M}_k}=\bigcup_{{p} \in {I}_k} \overline{M_p} $ be the union of
fibre spaces $\overline{M_p}$, ${p}\in {I}_k$ on ${\cal N}_k$, and
$\overline{F}({\overline{\cal M}_k})$ be the function space.

\omits{consider such a fibre bundle that the time $t$ as its base
space is
 discretized as
\begin{equation}\label{td} t\in R  \rightarrow
t\in  {T_D}=\{ (t_k , t_{k+1}=t_k+\Delta t_k, \quad k \in I)\},
\end{equation} but its $n$-dimensional fibre space $M_k$, at
moment $t_k (k \in I)$, is still continuous and smooth enough. Let
$N$ be the set of all nodes on ${T_D}$ with index set
$Ind({N})=I$, and ${M}=\bigcup_{k \in I} M_k$ be the fibre space
on $T_D$.
 At the moment $t_k$, ${\cal
N}_k$  denotes the set of nodes neighboring to $t_k$, and ${I}_k $
is the index set of nodes of ${\cal N}_k$ including $t_k$. The
coordinates of $M_k$ are denoted by $q^i({t_k})=q^{i (k)}, i=1,
\cdots, n$.  Let ${\cal M}_k=\bigcup_{{l} \in {I}_k} M_l $ be the
union of fibre spaces $M_l$, ${l}\in {I}_k$ on ${\cal N}_k$, and
$F({\cal M}_k)$ be the function space.}

Now we start from the case of $l=1$, where  Lagrangian function
depends on first order derivative only. In terms of  definition
2.1, the total exterior differential of $\Delta_k q^{i(k)}$ is
\begin{equation}\begin{array}{rcl}\label{deltaq}
d_t (\Delta_k q^{i(k)})&=&\Delta_k  (d_t q^{i(k)}) +
(\Delta_{k+1}q^{i(k+1)}-\Delta_{k}q^{i(k+1)}) \chi^E
 \\
  &=&\Delta_k  (d_v q^{i(k)})+
(\Delta_{k+1}q^{i(k+1)}
                 -\Delta_{k}q^{i(k)}) \chi^E \\
  &=&d_v (\Delta_k  q^{i(k)}) +[(R_E -id)\;\Delta_k  q^{i(k)}] \chi^E.
\end{array}
\end{equation}
This relation suggests that the  exterior differential  for
$v_1^{i(k)}$ should be as follows
\begin{equation}\begin{array}{rcl}
 d_t v_1^{i(k)}&:=& d_v v_1^{i(k)} + d_h v_1^{i(k)},\\
  d_h v_1^{i(k)}&:=&[(R_E -id)\;v_1^{i(k)}] \chi^E
     = (v_1^{i(k+1)}-v_1^{i(k)}) \chi^E.
\end{array}
\end{equation}
Here the action of translation operators $R_E$ and $R_f$ of
(\ref{rerf}) are extended to the function with the new coordinate
$ v_1^{i(k)}$  is :
\begin{equation}
R_E \; f(\{v_1\};\{t_k\}):= f(\{R_E\;v_1\};\{R_E \;t_k\}), \\
R_f \; f(\{v_1\};\{t_k\}):= f(\{R_E\;v_1\};\{t_k\}).
\end{equation}
where  $R_E v_1^{i(k)}:=v_1^{i(k+1)}$.

We can define the total exterior differential $d_t$ on the
 function which  depends on first order derivative only as:

\vspace{3mm}

{\it {\bf  Definition 2.2}:  the  NCDC on function $f\in
\overline{F}({\overline{\cal M}_k})$ can be defined :
\begin{equation}
d_t f(\{q\},\{v_1\};\{t_k\}) =\displaystyle\sum_{m=0}^{1}
\sum_{i,k\in\{v_m\}} (\partial_{v_m^{i(k)}} f)\; d_t v_m^{i(k)}
  +( R_E\; f - R_{f}\; f)\; \chi^E .
\end{equation} }


Similarly we can define NCDC on the case of function depending on
higher order derivatives as follows.

 \vspace{2mm}
 {\it {\bf  Definition 2.3}: If  $f\in
\overline{F}({\overline{\cal M}_k})$ the NCDC on function $f$ is:
\begin{equation}
  d_t f(\{q\},\{v_1\},\cdots,\{v_n\};\{t_k\})
 =\displaystyle\sum_{m=0}^{n} \sum_{i,k\in\{v_m\}}
  (\partial_{v_m^{i(k)}} f)\; d_t v_m^{i(k)} +( R_E\; f - R_{f}\; f)\;
  \chi^E ,
\end{equation}}
where
\begin{equation}\begin{array}{rcl}

 d_t v_m^{i(k)}&:=& d_v v_m^{i(k)} + d_h v_m^{i(k)},\\
 d_h v_m^{i(k)}&:=&[(R_E -id)\;v_m^{i(k)}] \chi^E
     = (v_m^{i(k+1)}-v_m^{i(k)}) \chi^E \\
     R_E f(\{v_m\},\{t_k\})&:=& f(\{R_E\;v_m\},\{R_E\;t_k\}),\\
R_f f(\{v_m\},\{t_k\}) &:= & f(\{R_E\;v_m\},\{t_k\}).
\end{array}\end{equation}

Then this operator $d_t$ on general functions case  is defined,
its nilpotency on the differential complex can be easily
constructed as in the last subsection. \omits{More details will be
published elsewhere.}


\section{Difference Discrete Variational Principle in Discrete Mechanics and Symplectic Algorithm}

In this section, we will study the variation problems in  discrete
mechanics including the symplectic algorithm completely based on
the established NCDC on the extended fibre bundle where the
functions may depend on high order derivatives. The method we will
use is the closest analogue of geometric-variational approach to
continuous mechanics. One of the key point is that the difference
as the discrete derivative in the NCDC is taken as independent
variable in the variation. Therefore, the variation of this kind
is called the difference discrete variational principle
\cite{glww,gw}.

\subsection{Difference Variation in Discrete Mechanics}
As were shown in \cite{gw}, total variation $\delta_t S$ in
continuous mechanics can be expressed as a contraction between
total exterior differentiation of action $S$ and  total
variational vector field $\xi_{total}$. Variational principle says
that it should be zero with fixed end points. On the other hand,
at the path which satisfies
Euler-Lagrange equation, 
the only contribution to $\delta_t S$ comes from the endpoint:
$\delta_t S= G(t_2)-G(t_1)$. This is the most general idea of
action principle in continuous mechanics.\omits{ \cite{DR94},} It
can  easily be generalized to the discrete mechanics.

When we discuss the variation problem in discrete mechanics using
 the extended fibre bundle with local bundle coordinates
 $(v_0^{i(k)}, v_1^{i(k)}, \cdots,v_l^{i(k)})$, it is
necessary to take  $v_m^{i(k)}=\Delta_k v_{m-1}^{i(k)}$,
$m=1,\cdots,l,$ and $q^{i(k)}=v_0^{i(k)}$.
But the exterior differentiation of
$f(v_m^{i(k)})$ is different from that of $f(\Delta_k
v_{m-1}^{i(k)})$ by the NCDC, expect for simple linear function.
One way to solve this problem is to introduce Lagrange multipliers
used at constrained dynamics, then we can take the discrete action
as:
\begin{equation}
S_D:= \displaystyle\sum_{k=1}^{N-1} \int_{D, t_k}^{t_{k+1}}
\{L(q^{i(k)},v_1^{i(k)},\dots, v_n^{i(k)};t_k)+
\sum_{m=1}^{l}\sum_{i} \lambda^{i(k)}_m (\Delta_k
v_{m-1}^{i(k)}-v_m^{i(k)}) \}\; dt_k,
\end{equation}
here $\lambda^{i(k)}_m$'s are the Lagrange multipliers with
dimension $(n\times l)$ and discrete integral $\int_{D}$ is
defined as in (\ref{itg}).
At the moment $t_k$, as the exterior
differentiation rules for each $\lambda^{i(k)}_m$ is taken  the
same form for $v^{i(k)}_m$, as in definition 2.3 except that
dimensions of each extended fibre space are doubled.

The variational vector field at moment $t_k$ is defined as
follows:
\begin{equation}
\xi_{total}^{(k)}:= \displaystyle
        \delta t_k \;\hat{\partial}_{t_k}
        + \sum_{m=0}^n\sum_{i,k'\in \{\cdot\}} (\delta_t v_m^{i(k')})\;
               \frac{\partial}{\partial v_m^{i(k')}}
        + \sum_{m=1}^n\sum_{i,k'\in \{\cdot\}} \delta_t\lambda^{i(k')}_m \;
               \frac{\partial}{\partial \lambda_m^{i(k')}}    ,
\end{equation}
where $\hat{\partial}_{t_k} g:=(t_{k+1}-t_k)^{-1} (R_E g-R_f g)$.
The variations $\delta t_k$, $\delta_t v_m^{i(k')}$ and$\delta_t
\lambda_m^{i(k')}$ can be defined from the
 contraction between
total exterior differentiation of $dt_k$, $d_tv_m^{i(k')}$ and
$d_t \lambda_m^{i(k')}$  and the basis of total variational vector
field $\xi_{total}$ as follows:
\begin{equation}\begin{array}{rcl}\label{contract}
\delta t_k &:=&
(t_{k+1}-t_k)\;<\chi^E,\;\xi_{total}^{(k)}>=<dt_k,\;
\xi_{total}^{(k)}>, \\
\delta_t v_m^{i(k')}
 &:=&<d_tv_m^{i(k')},\;\xi_{total}^{(k)}>
 =\delta_v v_m^{i(k')}+(\Delta_{k'} v_m^{i(k')}) \delta t_{k'},\;\;
 m=0,\dots,n, \\
\delta_t \lambda_m^{i(k')}
 &:=&<d_t \lambda_m^{i(k')},\;\xi_{total}^{(k)}>
 =\delta_v \lambda_m^{i(k')}+(\Delta_{k'} \lambda_m^{i(k')})
 \delta t_{k'},\;\;
 m=1,\dots,n.
\end{array}\end{equation}
\omits{------------------------- \vskip 2mm {\bf Remark 3.1:}
Although definition (\ref{base}) means the differential form
$dt_{k'}$, $k'\neq k$, is related to $dt_k$ by means of
$(t_{k'+1}-t_{k'})^{-1} dt_{k'} = (t_{k+1}-t_k)^{-1} dt_k$, we
must point out that the infinitesimal variable $\delta t_{k'}$ is
independent of $\delta t_k$. Now, the following problem is which
$\xi^{(k)}_{total}$ the induced differential form should contract
with? In discrete mechanics, a handy way is to express discrete
action as a (discrete) integral along (discrete) time coordinate,
since the definition of discrete integral (\ref{itg}) provides a
natural approach to determine the correct lattice node (or moment)
where the induced differential form is integrated.}

 Now, the variation of discrete action comes from two parts, one
is from contraction between $d_t \{L^{(k)}+\sum \lambda^{i(k)}_m
(\Delta_k v_{m-1}^{i(k)}-v_m^{i(k)})\}$ and $\xi_{total}^{(k)}$,
 another is from variation of the measure. At moment $t_k$, there
is $\delta (\int_{D, t_k}^{t_{k+1}} {\cal D} t_k)=\int_{D,
t_k}^{t_{k+1}} \delta ({\cal D} t_k)$. In comparison with equation
\begin{equation}
\delta (\int_{D, t_k}^{t_{k+1}} {\cal D} t_k) = \delta
(t_{k+1}-t_k)=\delta t_{k+1}-\delta t_k,
\end{equation}
it means
\begin{equation}
\delta ({\cal D} t_k) :=<d \;({\cal D} t_k),\;\xi_{total}^{(k)}>({\cal D} t_k) \; \Delta_k \delta t_k.
\end{equation}

\subsection{Difference Variation in Lagrangian Mechanics}

In discrete Lagrangian mechanics, a discrete version of
variational principle can be expressed as follows. At finite
interval $[t_1,t_N]$ (finite notes) of discrete time coordinate,
the only contribution to total variation of discrete action should
come from end point  term.

\vskip 3mm {\it Case 1:}  the $l$-order Lagrangian
$L^{(k)}=L(q^{i(k)},v_1^{i(k)},\dots, v_l^{i(k)};t_k)$:

The total exterior differential of $S_D^{(k)}$ is,
\begin{equation}
\begin{array}{rcl}
d_t S_D^{(k)} &=& \displaystyle \int_{D t_k}^{t_{k+1}}
\{(L^{(k+1)}+ \sum_{m=1}^{l} \sum_{i} \lambda^{i(k+1)}_m
(\Delta_{k+1}
v_{m-1}^{i(k+1)}-v_m^{i(k+1)})) \;d({\cal D}t_k)   \\
&&\quad + {\cal D}t_k\; d_t (L^{(k)}+ \sum_{m=1}^{l} \sum_{i}
\lambda^{i(k)}_m (\Delta_k v_{m-1}^{i(k)}-v_m^{i(k)}))\}.
\end{array} \end{equation}
After a straightforward computation we can get,
\begin{equation}
\begin{array}{rcl}\label{constraint}
d_t \; \{\lambda^{i(k)}_m (\Delta_k v_{m-1}^{i(k)})\} &=&
(\Delta_k v_{m-1}^{i(k)})\; d_t \lambda^{i(k)}_m +\lambda^{i(k)}_m
\Delta_k (d_t v_{m-1}^{i(k)})-\lambda^{i(k+1)}_m
(\Delta_{k+1} v_{m-1}^{i(k+1)})\Delta_k dt_k, \\
d_t \; \{\lambda^{i(k)}_m v_m^{i(k)})\} &=& v_m^{i(k)}\; d_t
\lambda^{i(k)}_m +\lambda^{i(k)}_m d_t v_{m}^{i(k)},
\end{array} \end{equation}
 Then the total variation of  $S_D^{(k)}$ becomes
\begin{equation}
\begin{array}{rcl}
\delta_t S_D^{(k)} &=&\displaystyle \int_{D t_k}^{t_{k+1}} {\cal
D}t_k\; \{\Delta_k (\sum_{m=1}^{l} \sum_{i} \lambda^{i(k-1)}_m
\delta_t v_{m-1}^{i(k)}-E^{(k)}\delta t_k)+(\Delta_k E^{(k)}
 +\hat{\partial}_{t_k} L^{(k)})\delta t_k \\
&& \;\;+\displaystyle\sum_i(\sum_{m=1}^{l} (\Delta_k
v_{m-1}^{i(k)}-v_m^{i(k)})\delta_t
\lambda^{i(k)}_m+(\partial_{q^{i(k)}}L^{(k)}-\Delta_k
\lambda^{i(k-1)}_1 )\; \delta_t q^{i(k)} \\
&& \;\; +\displaystyle
\sum_{m=1}^{l-1}(\partial_{v_m^{i(k)}}L^{(k)}-\Delta_k
\lambda^{i(k-1)}_{m+1}-\lambda^{i(k)}_{m})\;\delta_t
v_m^{i(k)}+(\partial_{v_n^{i(k)}}L^{(k)}- \lambda^{i(k)}_n )\;
\delta_t v_n^{i(k)} )\},
\end{array}
\end{equation}
where $E^{(k)}:=\sum_{m=1}^{l} (\sum_{i} \lambda^{i(k)}_m
v_{m}^{i(k)})-L^{(k)}$.

Variational principle says the bulk part of total variation must
be zero. After some algebraic computation we finally obtain,
\begin{equation}
\begin{array}{l}\label{n-order}
v_m^{i(k)}= \Delta_k v_{m-1}^{i(k)}, \quad\quad
\lambda^{i(k)}_{m}=\displaystyle(\sum_{h=0}^{l-m}(-\Delta_k
 R_E^{-1})^h \;\partial_{v_{m+h}^{i(k)}})\; L^{(k)}, \quad
 m=1,\dots,l, \\
\displaystyle \partial_{q^{i(k)}}
L^{(k)}+(\sum_{h=1}^{l}(-\Delta_k R_E^{-1})^h
\partial_{v_{h}^{i(k)}})\; L^{(k)}=0, \quad\quad
\Delta_k E^{(k)}
 +\hat{\partial}_{t_k} L^{(k)} =0,
\end{array}
\end{equation}

If the time step-lengths are fixed, the last equation of
(\ref{n-order}) has no solution in general even if the Lagrangian
does not depend on time manifestly.

\vskip 3mm {\it Case 2:} Lagrangian $L^{(k)}=L(\alpha
q^{i(k)}+\beta q^{i(k+1)},v_1^{i(k)};t_k)$: \vskip 2mm

Let $\partial_{q^{i}} L^{(k)}$ be the derivative w.r.t. variable
$\alpha q^{i(k)}+\beta q^{i(k+1)}$ in $L^{(k)}$, and
$\partial_{q^{i}} L^{(k-1)}$ be the derivative w.r.t. variable
$\alpha q^{i(k-1)}+\beta q^{i(k)}$ in $L^{(k-1)}$, etc. Denote
$\tau_k:=t_{k+1}-t_k$. By means of relation $\beta
\partial_{q^{i}}L^{(k)} \delta_t q^{i(k+1)}=\Delta_k (\tau_{k-1}
\beta
\partial_{q^{i}}L^{(k-1)} \delta_t
q^{i(k)})+\tau_k^{-1}\tau_{k-1}\beta \partial_{q^{i}}L^{(k-1)}
\delta_t q^{i(k)}$, the total variation of $S_D^{(k)}$ becomes
\begin{equation}
\begin{array}{rcl}\label{1-order}
\delta_t S_D^{(k)} & =&\displaystyle \delta_t \; \int_{D
t_k}^{t_{k+1}} {\cal D}t_k \{L^{(k)}+ \sum_{i} \lambda^{i(k)}
(\Delta_k q^{i(k)}-v_1^{i(k)}) \} \\
&=& \displaystyle \int_{D t_k}^{t_{k+1}}
 {\cal D}t_k \; \{ \Delta_k \Theta_D^{(k)}
 +(\Delta_k E^{(k)}+\hat{\partial}_{t_k} L^{(k)}) \delta t_k \\
&&\quad +\sum_i ((\Delta_k q^{i(k)}-v_1^{i(k)})\delta_t
\lambda^{i(k)} + [L_{q^{i(k)}}]\;\delta_t
q^{i(k)}+(\partial_{v_1^{i(k)}} L^{(k)}-\lambda^{i(k)})\delta_t
v_1^{i(k)}) \},
\end{array}\end{equation}
where
\begin{equation}
\begin{array}{rcl}
[L_{q^{i(k)}}]& := &\alpha
\partial_{q^{i}}L^{(k)}+\tau_k^{-1}\tau_{k-1}\;\beta
\partial_{q^{i}}L^{(k-1)}
 -\Delta_k(\lambda^{i(k-1)}),\\
E^{(k)}&:=& \sum_i \lambda^{i(k)}v_1^{i(k)}-L^{(k)}, \\
\Theta_D^{(k)}&:=& \sum_i (\lambda^{i(k-1)}+\tau_{k-1}
\beta\partial_{q^{i}}L^{(k-1)})\;\delta_t q^{i(k)}-E^{(k)} \delta
t_k.
\end{array}
\end{equation}

According to the discrete variational principle and the definition
of discrete integral (\ref{itg}), the discrete Euler-Lagrange
equations are,
\begin{equation}
\begin{array}{rcl}\label{motion}
v_1^{i(k)}=\Delta_k q^{i(k)},\quad && \lambda^{i(k)}\partial_{v_1^{i(k)}} L^{(k)},\\
\; [L_{q^{i(k)}}] = 0,  \quad && \Delta_k
E^{(k)}+\hat{\partial}_{t_k} L^{(k)} = 0.
\end{array} \end{equation}

Let $p_i^{(k)}:=\partial_{v_1^{i}} L^{(k-1)}+\tau_{k-1}
\beta\partial_{q^{i}}L^{(k-1)}$, so that  the variation equations
in (\ref{motion}) can be rewritten as follows:
\begin{equation}\label{ceq1}
\Delta_k p_i^{(k)}=(\alpha+\beta)
\partial_{q^{i}}L^{(k)}.
\end{equation}
Substitute $\partial_{q^{i}}L^{(k)}$ into
$p_i^{(k+1)}=\partial_{v_1^{i}} L^{(k)}+\tau_{k}
\beta\partial_{q^{i}}L^{(k)}$, we obtain
\begin{equation}\label{ceq2}
\frac{\alpha}{\alpha+\beta} p_i^{(k+1)}+\frac{\beta}{\alpha+\beta}
p_i^{(k)}=\partial_{v_1^{i}} L^{(k)}.
\end{equation}
It is just the discrete Legendre transformation, so that discrete
Hamiltonian at interval $[t_k,t_{k+1}]$ is naturally defined as
$H_D^{(k)}:= \sum_i (\frac{\alpha p_i^{(k+1)}+ \beta
p_i^{(k)}}{\alpha+\beta})v_1^{i(k)}-L^{(k)}$.

When $L^{(k)}=\frac{1}{2} (v_1^{i(k)})^2-V(\alpha q^{i(k)}+\beta
q^{i(k+1)})$ and $\alpha+\beta=1$, it is easy to verify that
equations (\ref{ceq1}) and (\ref{ceq2}) are just the discrete
canonical equations in Hamiltonian system $H=\frac{1}{2} {\bf
p}^2-V({\bf q})$. There are some interested cases: (i): $\alpha=0,
\beta=1$, (ii): $\alpha=1, \beta=0$, and
(iii):$\alpha=\beta=\frac{1}{2}$. Both case (i) and (ii) are the
known simple symplectic algorithms of order $1$ accuracy, and case
(iii) is the midpoint algorithms of order $2$.



Now, let's discuss constraint on $\alpha$ and $\beta$. For
simplicity  let $L^{(k)}=\frac{1}{2} (v_1^{i(k)})^2-V(\alpha
q^{i(k)}+\beta q^{i(k+1)})$, and $\tau_k\rightarrow 0,\; k\in Z$,
then we expand discrete Euler-Lagrange equations in $q^{i(k)}$
compare with Euler-Lagrange equation in
continuous mechanics, we have to take $\alpha+\beta=1$.

\subsection{Difference Variation in Hamiltonian Mechanics}

The total variation problem in the discrete Hamiltonian mechanics
with variable time step-lengths can be dealt with from the
viewpoint of NCDC too, where we should express  the discrete
Lagrangian in terms of the discrete Hamiltonian via Legendre
transformation.

The discrete version of variational principle is also expressed as
follows: at interval $[t_1,t_N]$ of discrete time coordinate the
total variation of discrete action comes from end point term only.
In Hamiltonian mechanics, if discrete Lagrangian is written as
$L=pv-H(p,q;t)$, then the  discrete action can be taken as
$\int_{D} {\cal D}t \;\{ pv-H(p,q;t)+\lambda (\Delta q-v)\} $.

\vskip 3mm

{\it Case 3:} $L^{(k)}=\sum_i (\alpha p_i^{(k+1)}+\beta
p_i^{(k)})v_1^{i (k)}- H^{(k)}$, where $\alpha+\beta=1$ and
$H^{(k)}:= H(\alpha p_i^{(k+1)}+\beta p_i^{(k)},\alpha
q^{i(k)}+\beta q^{i(k+1)};t_k)$: \vskip 3mm

Introduce Lagrange multipliers $\lambda^{i(k)}$ as in case 2. Let
$H_{p_i}^{(k)}$ denote derivative w.r.t. variable $\alpha
p_i^{(k+1)}+\beta p_i^{(k)}$ in $H^{(k)}$, and $H_{q^i}^{(k)}$
denote derivative w.r.t.  variable $\alpha q^{i(k)}+\beta
q^{i(k+1)}$ in $H^{(k)}$ respectively. Its total variation of
action at interval $[t_1,t_k]$ reads
\begin{equation}\label{tvdsh}
\begin{array}{rcl}
\delta_t S_D^{(k)}&=&\displaystyle \int_{D t_k}^{t_{k+1}} {\cal
D}t_k \; \{\Delta_k \Theta_D^{(k)}
 +(\Delta_k E_D^{(k)} -\hat\partial_{t_k} H^{(k)}) \delta t_k \\
&&\quad + \sum_i ((\alpha p_i^{(k+1)}+\beta
p_i^{(k)}-\lambda^{i(k)})\delta_t v_1^{i (k)}+
(\Delta_k q^{i(k)}-v_1^{i(k)})\delta_t \lambda^{i(k)} \\
 &&\quad \;\;+  (\beta(v_1^{i (k)}-
H_{p_i}^{(k)})+\alpha \tau_k^{-1}\tau_{k-1}(v_1^{i(k-1)}-
H_{p_i}^{(k-1)}))\;
\delta_t p_i^{(k)}\\
&&\quad  \;\; - (\Delta_k \lambda^{i(k-1)}+\alpha
H_{q^i}^{(k)}+\beta \tau_k^{-1}\tau_{k-1}H_{q^i}^{(k-1)}) \delta_t
q^{i(k)}) \},
\end{array}\end{equation}
where
\begin{equation}\label{energy}
\begin{array}{rcl}
\Theta_D^{(k)}&:=&\sum_i (\lambda^{i(k-1)}-
 \beta\tau_{k-1}H_{q^i}^{(k-1)}) \delta_t q^{i (k)} \\
 &&+\alpha \tau_{k-1}\; \sum_i(v_1^{i(k-1)}- H_{p_i}^{(k-1)})\;
 \delta_t p_i^{(k)}- E_D^{(k)} \delta t_k, \\
E_D^{(k)}&:=&\sum_i \lambda^{i(k)}v_1^{i(k)}-L^{(k)}.
\end{array}\end{equation}
Thus we obtain
\begin{equation}\label{cneqd1}
v_1^{i(k)}=\Delta_k q^{i(k)}, \quad\quad \lambda^{i(k)}=\alpha
p_i^{(k+1)}+\beta p_i^{(k)},
\end{equation}
 and discrete  equations of motion,
\begin{equation}\label{cneqd2}
\begin{array}{l}
\beta(v_1^{i (k)}- H_{p_i}^{(k)})+\alpha
\tau_k^{-1}\tau_{k-1}(v_1^{i(k-1)}- H_{p_i}^{(k-1)})=0, \\
  \Delta_k \lambda^{i(k-1)}
 +\beta \tau_k^{-1}\tau_{k-1}H_{q^i}^{(k-1)}+\alpha
 H_{q^i}^{(k)}=0,
\end{array}
\end{equation}
and the equation for the variable time step-lengths,
\begin{equation}\label{timesteph}
\Delta_k E_D^{(k)} -\hat\partial_{t_k} H^{(k)}=0.
\end{equation}
Which are nothing but the exact symplectic energy-conserving
algorithms.

In terms of relation (\ref{energy}),(\ref{cneqd2}) there is
$E_D^{(k)}=H^{(k)}$. Moreover, the discrete  equations of motion
(\ref{cneqd2}) become
\begin{equation}
\begin{array}{l}
\beta\tau_{k}(\Delta_k q^{i (k)}- H_{p_i}^{(k)})+\alpha
 \tau_{k-1}(\Delta_{k-1} q^{i (k-1)}-
 H_{p_i}^{(k-1)}))=0, \\
\alpha\tau_{k}(\Delta_k p_i^{ (k)}+ H_{q^i}^{(k)})+\beta
 \tau_{k-1}(\Delta_{k-1} p_i^{(k-1)}+
 H_{q^i}^{(k-1)})) =0.
\end{array}
\end{equation}
It should be satisfied at all nodes on lattice $T_D$, so that we
have to take
\begin{equation}\label{cneqd3}
\Delta_k q^{i (k)}- H_{p_i}^{(k)}=0, \quad\quad \Delta_k p_i^{
(k)}+ H_{q^i}^{(k)}=0.
\end{equation}

In this way, we have discrete Legendre transformation $v_1^{i(k)}-
H_{p_i}^{(k)}=0$ that transfer discrete Hamiltonian to discrete
Lagrangian on fibre bundle. And the end point terms become
standard form $\Theta_D^{(k)}=\sum_{i} p_i^{(k)} \delta_t q
^{i(k)} -H^{(k)} \delta t_k$. This case can be regarded as the
counterpart (in Hamiltonian formulism) of case 2.

>From $\delta_v S_D$ part in (\ref{tvdsh}), it is easy to see that
if we only take $d_v$ on $S_D$, it also can be divided into two
parts:
\begin{equation}\label{dSD}
d_v S_D=\displaystyle  \sum_k \int_{D,t_k}^{t_{k+1}}
 {\cal D}t_k\; \{\Delta_k {\theta_D}^{(k)} +{{\cal
E}_D}^{(k)} \},
\end{equation}
here ${{\cal E}_D}^{(k)}$, ${\theta_D}^{(k)}$ are the discrete
Euler-Lagrange 1-form and symplectic potential 1-form
respectively. Due to the nilpotency of $d_v$, it is
straightforward to get
\begin{equation}\label{ddsD}
d_v{{\cal E}_D}^{(k)}+\Delta_k {\omega_D}^{(k)}=0.
\end{equation}
Here differential structure ${\omega_D}^{(k)}:d_v{\theta_D}^{(k)}$ is just the 
 symplectic structure $\sum_i d_v p_i^{(k)}\wedge d_vq^{i (k)}$.
Therefore, we may get the discrete version for the theorem of
Euler-Lagrange cohomology and symplectic structure-preserving
condition \cite{glw}, \cite{glww}:

\vskip 3mm {\it Theorem: For a discrete Hamiltonian system with
$H(p,q;t)$, we have:

1. The discrete version of the first Euler-Lagrange cohomology may
be nontrivial: {\centerline{$H^{(1)}_{DCM}$:=\{Closed
Euler-Lagrange forms\}/ \{Exact Euler-Lagrange forms\}.}}

 2. The necessary and
sufficient condition for conservation of the discrete symplectic
2-form, i.e.
\begin{equation}\label{csvsyD}
\Delta_k{\omega_D}^{(k)}=0,
\end{equation}
is the corresponding discrete Euler-Lagrange 1-form being closed.
}

\section{Solution Existence and Site Density}

In the discrete evolution problem, if there has solution whose
discrete time series satisfy $ \tau_{k+1} =\tau_k+\delta
\tau_k=\tau_k +O(\tau_k^2 )$, we say there has a stable solution.
In this section, we discuss the case of $H(p,q;t)= {1\over 2}p^2
+V(q,t) $ with midpoint algorithms, and find this algorithm will
limit the form of $V(q,t)$ in order that discrete system have
stable solution.

Now, let us analyze the discrete evolution problem of
$t_k\rightarrow t_{k+1} $ and $t_{k+1}\rightarrow t_{k+2}$. In
these two steps, we have four discrete canonical equations,
\begin{equation}
\begin{array}{cc}
\displaystyle {p^{(k+1)} -p^{(k)} \over t_{k+1}-t_k} =- V^\prime (
{q^{(k+1)}+q^{(k)} \over 2}, {t_{k+1}+t_k \over 2} ) , \;&
\displaystyle {q^{(k+1)} -q^{(k)} \over t_{k+1}-t_k}= {p^{(k+1)}+p^{(k)} \over 2}, \\
\displaystyle {p^{(k+2)} -p^{(k+1)} \over t_{k+2}-t_{k+1}} =-
V^\prime ( {q^{(k+2)}+q^{(k+1)} \over 2}, {t_{k+2}+t_{k+1} \over
2} ) ,\;& \displaystyle {q^{(k+2)} -q^{(k+1)} \over
t_{k+2}-t_{k+1}}= {p^{(k+2)}+p^{(k+1)} \over 2},
\end{array}
\end{equation}
and two discrete energy equations, one of which is
\begin{equation}
\begin{array}{rcl}
&& \displaystyle \frac{1}{2} (\frac{p^{(k+1)}+p^{(k)}}{2} )^2
+V(\frac{q^{(k+1)}+q^{(k)}}{2}, \frac{t_{k+1}+t_k }{2} ) \\
 &=&\displaystyle \frac{1}{2}  (\frac{p^{(k+2)}+p^{(k+1)}}{2} )^2
+V(\frac{q^{(k+2)}+q^{(k+1)}}{2}, \frac{t_{k+1}+t_k}{2} ).
\end{array}
\end{equation}

Above five equations are enough to discuss the relation of
$\tau_{k}$ and $\delta\tau_{k}$ as follows. Firstly, let's denote
\begin{equation}
\begin{array}{cc}
t_{k+1}-t_k =\tau_k,& \quad t_{k+2}-t_{k+1} =\tau_k + \delta \tau_k , \\
p^{(k+1)} -p^{(k)} =r , &\quad  p^{(k+2)} -p^{(k+1)} = r+ \delta r ,\\
q^{(k+1)} -q^{(k)} =s , &\quad q^{(k+2)}-q^{(k+1)} = s + \delta s.
\end{array}
\end{equation}
In this way, the quantities on the $k$-th and $(k+2)$-th nodes can
be expressed by those quantities on the $(k+1)$-th node and
$\delta \tau_k$, $\delta r$ and $\delta s$. At the same time, we
also use notation $p=p^{(k+1)}$, $ q=q^{(k+1)}$ and $t=t_{k+1} $
for simplicity, so those five equations become
\begin{equation}
\begin{array}{rcl}
\displaystyle {r \over \tau_k} =- V^\prime  ( q- {s \over 2} ,t
-{\tau_k \over 2} ), &&  \quad\quad\quad
\displaystyle {s\over \tau_k} =(p - {r \over 2} ),  \\
\displaystyle {r + \delta r \over \tau_k + \delta \tau_k} =- V
^\prime ( q+ {s+ \delta s \over 2}, t+ {\tau_k+ \delta \tau_k
\over 2} ), && \quad\quad\quad
\displaystyle {s+ \delta s \over \tau_k + \delta \tau_k} = ( p + {r + \delta r \over 2} ), \\
\displaystyle {1 \over 2} (p-{r \over  2} )^2 + V(  q- {s \over 2}
,t -{\tau_k \over 2} ) &=& \displaystyle {1\over  2} ( p + {r +
\delta r \over 2} ) ^2 + V( q+ {s+ \delta s \over 2} ,t -{\tau_k
\over 2} ).
\end{array}
\end{equation}
Secondly, If $\tau_k$ is a small time step-length and there are
$r,s \sim\tau_k$ and $\delta r,\delta s,\delta \tau_k \sim
\tau_k^2$, we can expand above equations in Tailor Series and omit
its higher order terms than $\tau_k^2$. The new five equations can
be regarded as constraint equations of six variables $\tau_k$,
$r$, $s$, $\delta \tau_k$, $\delta r$ and $\delta s$, so we can
finally obtain an equation of $\tau_k$ and $\delta \tau_k$:
\begin{equation}\label{cl0}
V^\prime_t p\; \tau_k +\frac{1}{2}( V^{\prime \prime} p^2  +
3V^\prime _t p +  V^{\prime 2})\; \delta \tau_k + \frac{1}{2}(
V^{\prime \prime } _t p^2 + \frac{1}{3}V^{\prime \prime \prime
}p^3)\; \tau_k^2 = 0 .
\end{equation}

If the perturbation term gives us a small parameter $\sim$
$V_t/V$, which is less than or comparable with a finite small
$\tau_k$, it should have a solution in the domain $V^{\prime
\prime} p^2 + 3V^\prime _t p + V^{\prime 2}\neq 0$, as showed in
the next section. In the other words, hypothesis $\delta \tau_k
\sim \tau_k^2$ should be modified to $\delta \tau_k \sim
\varepsilon(t_k) \tau_k$, where $\varepsilon(t_k)$ is a small
quantity but not tend to zero in continuous limit. But in the case
of finite small $\tau_k$, we can keep the form of (\ref{cl0})
since $\varepsilon(t_k) \tau_k$ is comparable with the $2$-order
term $\tau_k^2$ or its higher order term in Tailor series.

\vskip 3mm

If Hamiltonian does not depend on time manifestly, the hypothesis
$\delta \tau_k \sim \tau_k^2$ can be kept even in the continuous
limit. Now, (\ref{cl0}) becomes
\begin{equation}\label{cl}
(3 V^{\prime \prime} p^2  + 3 V^{\prime 2})\; \delta \tau_k +
V^{\prime \prime \prime }p^3 \; \tau_k^2 =0 ,\quad \hbox{or} \quad
\frac{\delta \tau_k}{\tau_k}=- \frac{V^{\prime \prime \prime
}p^3}{(3 V^{\prime \prime} p^2 + 3 V^{\prime 2}) }\; \tau_k.
\end{equation}
It means ${\delta \tau_k}/{\tau_k} \rightarrow 0$ too, so we have
\begin{equation}
 \frac{\tau_{k+N}}{\tau_k}=\prod_{i=0}^{N-1}
  \frac{\tau_{k+i+1}}{\tau_{k+i}}=\prod_{i=0}^{N-1}
  (1+\frac{\delta \tau_{k+i}}{\tau_{k+i}})\approx\prod_{i=0}^{N-1}
  e^{\frac{\delta \tau_{k+i}}{\tau_{k+i}}}=e^{\sum_{i=0}^{N-1}
  \frac{\delta \tau_{k+i}}{\tau_{k+i}}}.
\end{equation}
According to equation (\ref{cl}) and definition of Riemann
integral, above equation can be rewritten in integral form. In
this way, discrete mechanics in continuous limit will give us more
information than usual continuous mechanics \cite{td2}. For
example, if we define site density at $t=t_i$ be
$\rho(t_i)=1/\tau_i$, above equation means the relation of site
densities between time $t_A$ and $t_B$ is
\begin{equation}
 \ln \frac{\rho(t_A)} {\rho(t_B)}=\int_{t_B}^{t_A} \frac{V^{\prime \prime \prime
}p^3}{(3 V^{\prime \prime} p^2 + 3 V^{\prime 2}) }
dt=\frac{1}{3}\ln (V^{\prime \prime} p^2 + V^{\prime
2})\Big|_{t_B}^{t_A}.
\end{equation}
Here continuous canonical equations are used in integrating. It
should marked that it needs $ V^{\prime \prime} p^2 + V^{\prime
2}\neq 0$ in $[t_A,\; t_B]$, or else higher order terms in Tailer
series of algorithms are needed.

It is obvious that $\rho(t)=\hbox{constant}$ for $V(q)=a q^2 +b
q$, since there is $V^{\prime\prime\prime}=0$.

\section{The Energy-Conserving Midpoint Scheme and Numerical Examples}
In this section, we apply the energy-conserving midpoint scheme to
Hamiltonian system that may depend on time manifestly.

Here we consider a pendulum with periodic perturbation, whose
continuous Hamiltonian is $H(l,\theta;t)=\frac{1}{2 M r^2} l^2 +M
g r (1-\cos \theta)*(1-0.1 * \sin(0.02 *t))$ where angular
momentum $l$ and angular coordinate $\theta$ are the canonical
variables. Take its discrete Hamiltonian as $H^{(k)}:H(\frac{l^{(k)}+l^{(k+1)}}{2},\frac{\theta^{(k)}+\theta^{(k+1)}}{2};
\frac{t_k+t_{k+1}} {2})$, so that energy equation
(\ref{timesteph}) becomes
$H(\frac{l^{(k)}+l^{(k+1)}}{2},\frac{\theta^{(k)}+\theta^{(k+1)}}{2};
\frac{t_k+t_{k+1}}{2})
=H(\frac{l^{(k+1)}+l^{(k+2)}}{2},\frac{\theta^{(k+1)}+\theta^{(k+2)}}{2};
\frac{t_k+t_{k+1}} {2}).$

If $M r^2=1 \; kg\cdot m^2$, $M g r=1 \; kg\cdot m^2/s^2$ and
initial values are $l^{(1)}=0.5 \; kg\cdot m^2/s$,
$\theta^{(1)}=0.5$, $t_1=0 s$ and $\tau_1=0.5 s$, then the time
step-length $\tau_k:=t_{k+1}-t_k$ varies 
according to the energy preserving equation (\ref{timesteph}). The
numerical results   are shown in the following three figures.

 Fig. 1 is a plot of time step-length $\tau_k$ versus time $t_k$
in $0\leq t_k \leq 10000 s$. It shows there is a long period about
${{2\pi}\over 0.02} \; s$ that comes from perturbation $-0.1
* \sin(0.02 *t)$.

Fig. 2 shows its detail in $0\leq t_k \leq 350 s$. Besides the
long period ${{2\pi}\over 0.02}\; s$, there is another short
period mainly from the system without considering perturbation.

Fig. 3 shows the detail in $0\leq t_k \leq 16 s$. The upper graph
shows angular coordinate $\theta^{(k)}$ versus time $t_k$, and the
lower graph is $\tau_k$ versus $t_k$. From these two graph, we
find it gives a relative smallest time step-length when
$\theta^{(k)}$ is near zero, and the relative biggest time
step-length appears when $\theta^{(k)}$ is near an extremum.

\begin{figure}
\begin{center}
\psfig{file=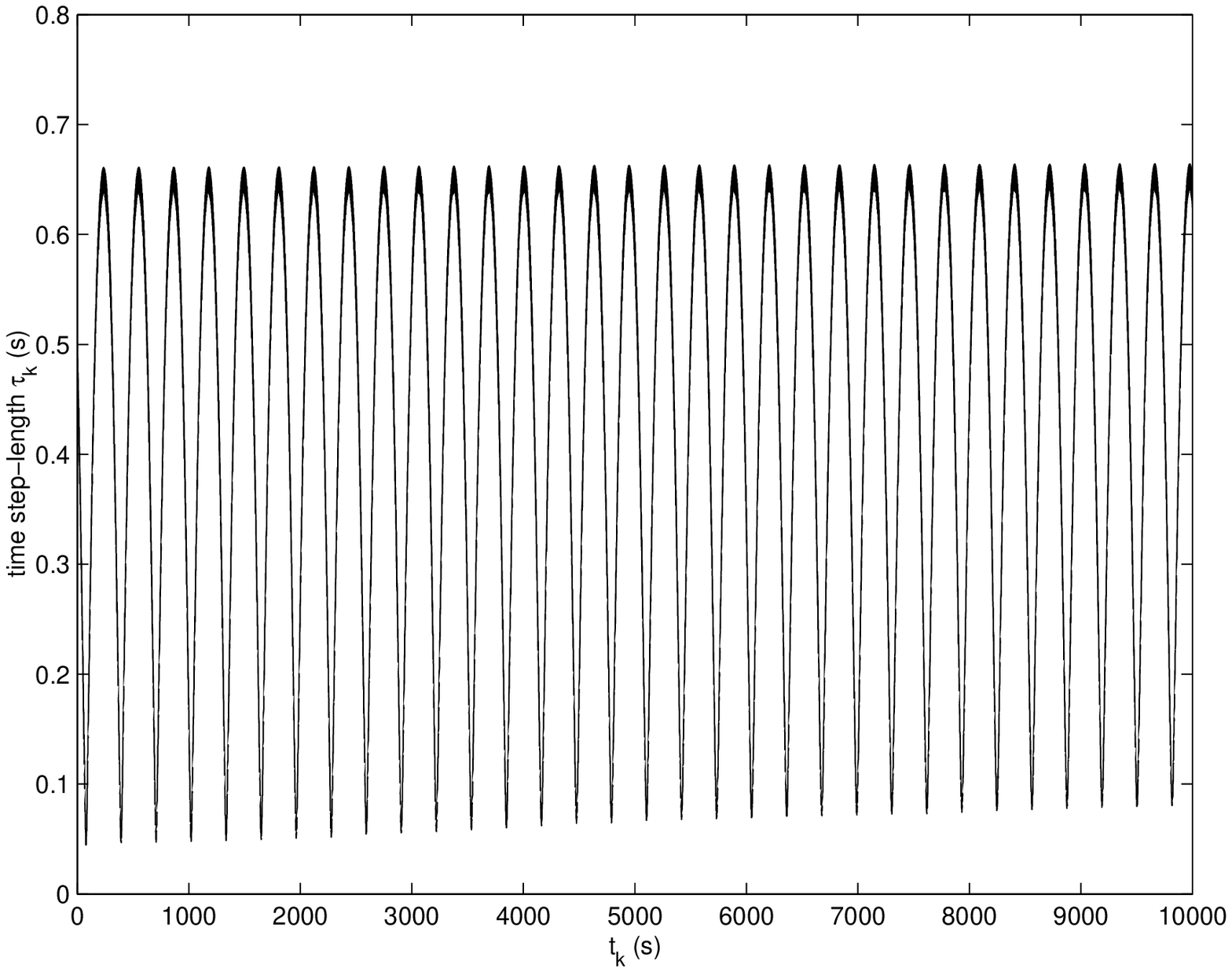,width=12cm} \caption{Time step-length $\tau_k$
versus time $t_k$, $0\leq t_k \leq 10000 s$.} \label{figl}
\end{center}
\end{figure}

\begin{figure}
\begin{center}
\psfig{file=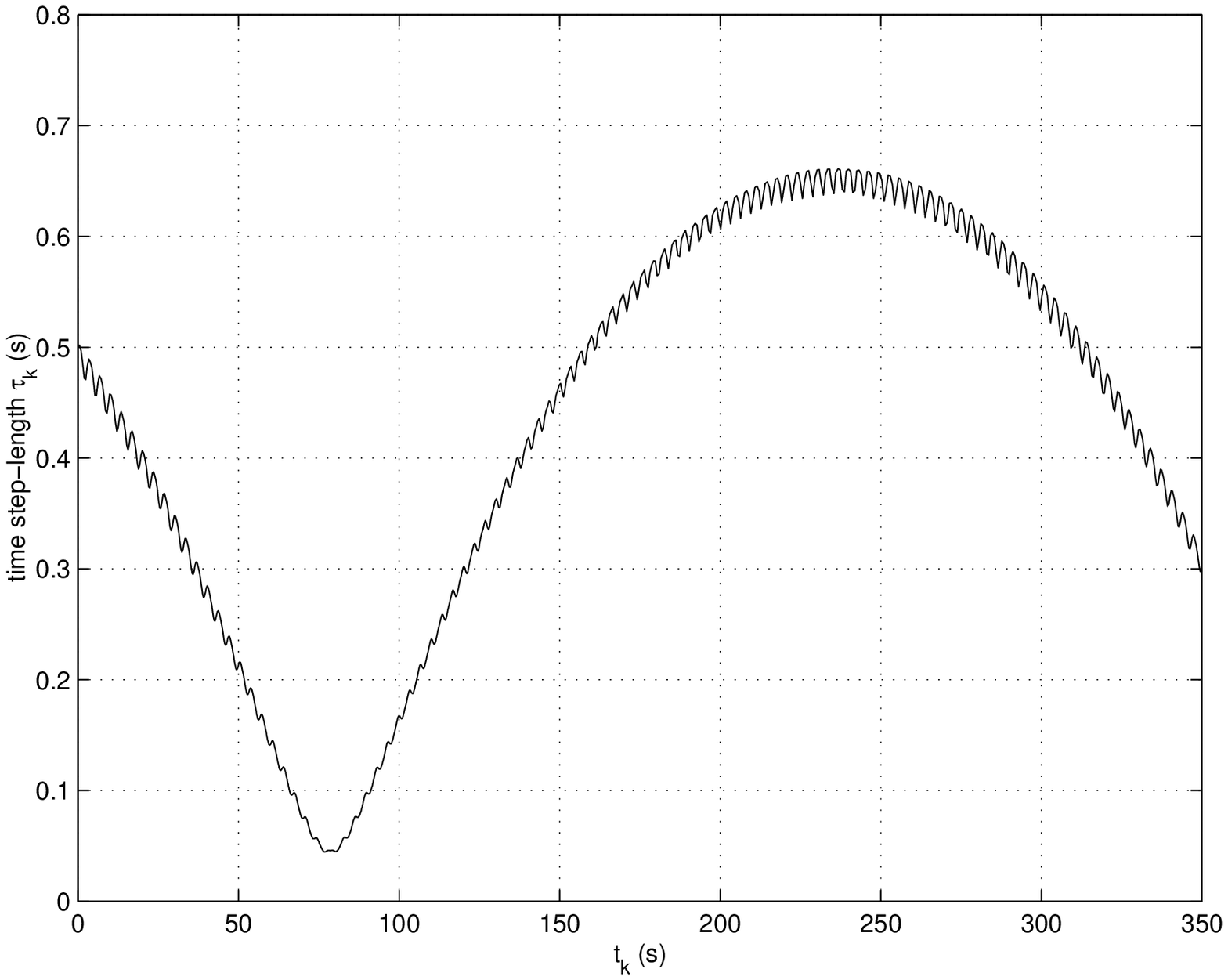,width=12cm} \caption{Time step-length $\tau_k$
versus time $t_k$, $0\leq t_k \leq 350 s$.} \label{fig2}
\end{center}
\end{figure}

\begin{figure}
\begin{center}
\psfig{file=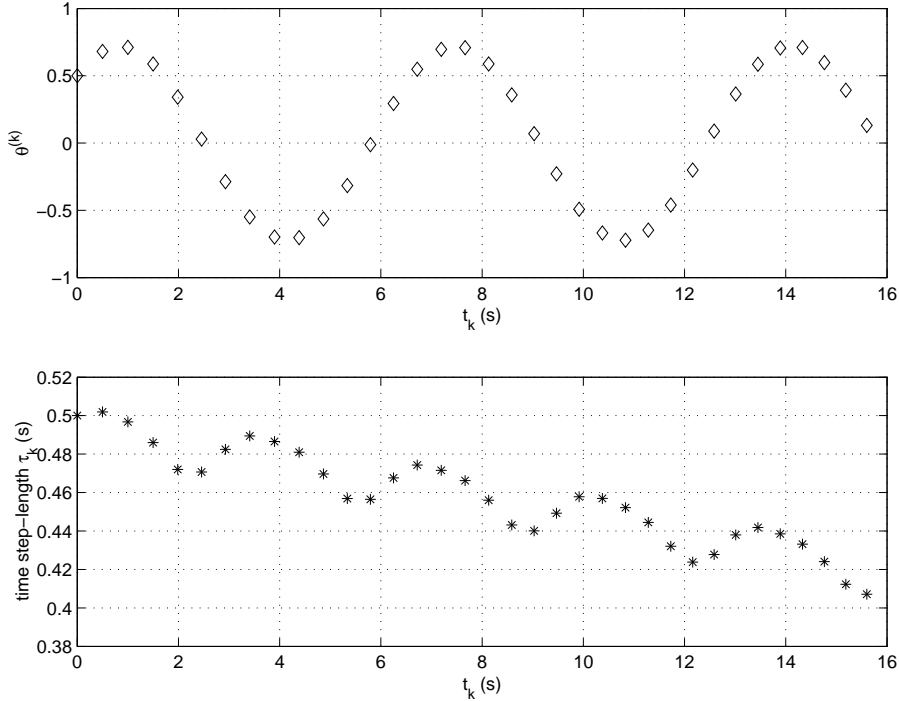,width=12cm} \caption{ $\theta^{(k)}$ and
$\tau_k$ versus $t_k$, $0\leq t_k \leq 16 s$. } \label{fig3}
\end{center}
\end{figure}

\section{Concluding Remarks}

In this paper, we establish an NCDC on 1-dimensional lattice with
variable step-lengths. Based upon this NCDC, we propose the
difference discrete variational principle in discrete mechanics
for the finite time interval  and apply it to both Lagrangian and
Hamiltonian formalisms with higher order derivative with respect
to discrete time. The two formalisms are related by discrete
Legendre transformation. In order to deal the cases with higher
derivatives, we make use of the Lagrange multiplier\omits{, which
is different from \cite{gw}}.
Moreover we discuss some the discrete mechanics and symlectic
algorithm in both Lagrangian and Hamiltonian formalism. We obtain
corresponding symplectic and energy-preserving schemes and we also
show that the necessary and sufficient conditions for the
symplectic 2-form preserving in discrete mechanics are the
corresponding Euler-Lagrange 1-form being closed \cite{gw,glw}.

When Lagrangian/Hamiltonian depend on time manifestly, we find
that the discrete energy conservation equation  in our approach is
different  from Lee's approach \cite{td1,td2,td3} or Veselov's
generalized version \cite{MPS98}. The reason comes from whether
the NCDC in base space as well as in fiber bundle is suitably
considered.
In our approach, variation problems are completely dealt with from
the viewpoint of NCDC, which is the closest analogous of
continuous mechanics in the view of geometric-variational
approach.

Moreover, we show that the discrete mechanics in continuous limit
may give some information of site density that cannot be obtained
in usual continuous mechanics \cite{td2}. Here we also find that
there is $\ln \rho(t)\sim \frac{1}{3}\ln (V^{\prime \prime} p^2 +
V^{\prime 2})$ when Hamiltonian does not depend on time
manifestly.

Finally, it should be mentioned that as what had been done in
\cite{gw} our approach with the NCDC may be generalized to
discrete classical field theory and the multisymplectic algorithm
with discrete energy-momentum preserving. We will leave this
subject for further publication.

 \vskip 8mm

\centerline{\bf Acknowledgments} \vskip 2mm

This work was supported in part by the National Natural Science
Foundation of China (grant Nos. 90103004, 10171096) and the
National Key Project  for Basic Research of China (G1998030601).

\end{document}